**Integrating Computational Methods and AI into Qualitative Studies of Aging and Later Life[1]**

Corey M. Abramson, Ph.D.

Associate Professor of Sociology
Center for Computational Insights on Inequality and Society (CIISR)
Rice University

**Abstract**
This chapter demonstrates how computational social science (CSS) tools are extending and expanding research on aging. The depth and context from traditionally qualitative methods such as participant observation, in-depth interviews, and historical documents are increasingly employed alongside scalable data management, computational text analysis, and open-science practices. Machine learning (ML) and natural language processing (NLP), provide resources to aggregate and systematically index large volumes of qualitative data, identify patterns, and maintain clear links to in-depth accounts. Drawing on case studies of projects that examine later life--including examples with original data from the DISCERN[2] study (a team-based ethnography of life with dementia) and secondary analyses of the American Voices Project (nationally representative interview)--the chapter highlights both uses and challenges of bringing CSS tools into more meaningful dialogue with qualitative aging research. The chapter argues such work has potential for (1) streamlining and augmenting existing workflows, (2) scaling up samples and projects, and (3) generating multi-method approaches to address important questions in new ways, before turning to practices useful for individuals and teams seeking to understand current possibilities or refine their workflow processes. The chapter concludes that current developments are not without peril, but offer potential for new insights into aging and the life course by broadening--rather than replacing-- the methodological foundations of qualitative research.

---

[1]This research was supported by National Institute on Aging of the National Institutes of Health (NIA/NIH) award DP1AG069809 (Dohan PI). Content and views are those of the author not of NIH. I would like to extend gratitude to Daniel Dohan, Alissa Sideman, Alma Hernandez, Ignacia Arteaga, Melissa Ma, Brandi Ginn, and Yuhan (Victoria) Nian for related work, the American Voices Project (AVP) team for their support in working with that large-scale qualitative data, the Qualitative Data Repository at Syracuse University for supporting efforts to verify associated papers, the Russell Sage Foundation for facilitating opportunities to interact and share thoughts on open qualitative social science, and the members of the National Dementia Workforce Study for thoughtful discussions (particularly Spetz, Maust, Yeh, Arteaga, and Bates).

[2] NIH NIA DP1 AG069809 (PI: Dohan), Title: Next-Gen ethnography to understand decision-making among diverse populations impacted by Alzheimer's disease and related dementias (ADRD). DISCERN, derived from: "DIverSe Cultures, EthnogRaphy, and decision making in demeNtia — implications for Medical Culture"

## 1. Introduction

Qualitative research methods--especially participant observation and in-depth interviewing--have long been central to social scientific understandings of aging. By immersing themselves in the communities, institutions, and cultural contexts where older adults live, qualitative researchers highlight processes and experiences that broader survey methods might miss or only partially capture (Gubrium, 1997; Hochschild, 1973; Kaufman, 1986; Loe, 2011; Myerhoff, 1980; Newman, 2004). From Myerhoff's (1980) depiction of Jewish immigrant identities, to Hochschild's (1973) observation of how social engagement can be maintained among older adults, and Gubrium's (1976/1997) analysis of institutional life, participant observation and in-depth interviews reveal everyday life worlds, interactions, and repertoires that shape possibilities in everyday life. This provides a necessary vantage to understand how older adults navigate complex and often unequal social structures with granularity a single method cannot capture, as well as how institutions shape the lives of older adults (Abramson, 2015; Ferraro, 2018; Torres, 2025).

Although qualitative methods offer the possibility of rich, contextual understandings central to sociological inquiry (Cicourel, 1982), they have historically faced challenges around scope, transparency, replicability, and scalability (Abramson & Dohan, 2015; Goldthorpe, 2000; Murphy et al., 2021). Yet, the landscape of qualitative research is changing. In studies of aging, ethnographic approaches are increasingly a steady fixture, in journal articles and competitive grants as well as scholarly monographs (Abramson, 2021; Phoenix, 2018). Meanwhile, ongoing transformations in digital technology--including artificial intelligence (AI) and computational social science--present new avenues for analyzing data at scale, including information in the form of qualitative text like transcripts and fieldnotes, alongside new challenges (Burrell & Fourcade, 2021; Farber et al., 2025; Li & Abramson, 2025; Nelson, 2020; Roberts et al., 2022). Researchers increasingly ask whether the interpretive depth of qualitative methods can be maintained--or possibly enriched--by contemporary computational tools, machine-assisted workflows, and larger qualitative data sets (Abramson et al., 2018, 2026; Brower et al., 2019; Chandrasekar et al., 2024; Small, 2011). These questions echo broader shifts in sociology, where computational social science (CSS) analyses of "big data" intersect with traditional methodological approaches (Breiger, 2015; Nelson et al., 2021).

An increasing body of literature identifies both pathways and alternatives to traditional approaches in an era of increasing computation (Abramson et al., 2026; Bail, 2024; Edelmann et al., 2020). In a recent *Annual Review of Sociology* piece, following the twenty-fifth anniversary of the classical debates on the use of computers in systematic qualitative analysis, my colleagues and I argued that computational--qualitative engagements often mirror a typology of practices that either extend or reconfigure qualitative boundaries by working to-- (1) streamline and augment existing workflows, (2) scale up samples and projects, (3) generate hybrid analytical approaches, and (4) interrogate the sociology of computation itself, both as a topic and act of self-reflection (Abramson et al., 2026). These uses connect directly to growing works examining questions of aging-- both in our own studies and in national initiatives. For example, the American Voices Project (AVP) shares large-scale in-depth interview data with teams working on projects using "big qualitative data" to reveal health inequalities across the life course in ways not possible in smaller-scale qualitative work (Abramson et al., 2024). Meanwhile, the National Dementia Workforce Study (NDWS) offers a new example that links survey data to computationally enhanced qualitative approaches to understand the care economy and workforce challenges in dementia care within and across occupational groups (Maust et al., 2025). Team-based ethnographic studies likewise use computational analyses to validate patterns and typologies (Arteaga et al., 2025). New data sets for science and policy revisit the possibility of public use portals with deidentified qualitative data for expanding the scale and accessibility of qualitative text (Abramson & Dohan, 2015; Edin et al., 2024; Gupta et al., 2021; Mauldin et al., 2024). Taken together, such cases show how the intertwined development of computational and qualitative projects can deepen our capacity to understand complex phenomena such as human aging.

Participant observation and in-depth interviews conducted in classic ways remain indispensable for understanding older adults' lives, experiences, and trajectories. Computationally assisted approaches conducted at scale, involve different study designs, and are not a direct analog or replacement (Abramson & Dohan, 2015; Grigoropoulou & Small, 2022). Yet the tools used in these approaches--ranging from supervised machine learning to identify topics in text, to large scale stratified sampling to better generalize from in depth interviews, to text mining for charting patterns in language--each offer avenues of extension and expansion for analyzing large volumes of qualitative data. They likewise extend work on cross-site comparisons, pattern visualization, and triangulating analyses in ways not feasible without computation (Abramson et al., 2018; Brower et al., 2019; Edin et al., 2024). While some of these techniques would correctly be labeled 'AI' in that they use advanced computation to look at language clusters (semantic networks), train a sentence-transformer model to classify mentions of dementia (machine learning), or help deidentify transcripts for a repository (named entity recognition), the discussion is grounded in scientific uses and aims to be precise. The core position is that when used purposefully, computational tools can support rather than replace interpretive expertise and expand capacity for understanding aging-- extending a longstanding tradition of pragmatically mixing methods to examine an array of sociological questions, topics and theories of interest (Abramson et al., 2024; Small, 2011).

By examining both the possibilities and pitfalls of contemporary scholarship, this chapter highlights how computational tools are being used to scale, rather than outsource, human interpretation. Rather than displacing field studies or in-depth interviews, the integration of computational tools may reduce the manual burden of large-scale coding and deidentification or help uncover patterns that might otherwise remain hidden. Robust field engagement, in turn, ensures that automated classifications remain grounded in real-world contexts, preventing superficial or misleading analyses associated with mistaking numeric scope for correspondence (McFarland & McFarland, 2015). This synergy between field notes, interview data, and computational text analytics paves new pathways for understanding how older adults navigate health systems, economic constraints, family networks, and broader cultural identities in later life (Arteaga et al., 2025; Li & Abramson, 2024), though--as with any method--its misuse can reproduce erroneous results.

This chapter proceeds as follows. First, the text briefly reviews contributions of qualitative work on aging, showing some key ways participant observation, in-depth interviews, and document analyses have historically advanced understandings of aging and the life course. Next, the chapter examines the expansion of computational approaches in qualitative data analysis, illustrating how large-scale text analysis and machine learning can complement interpretive depth. The chapter then turns to case studies that show examples of integrations of computational tools with qualitative methods, highlighting benefits and logistical challenges of both use for new projects and secondary analysis of archived qualitative data. Each is followed by a discussion of practical workflows for managing, analyzing, and interpreting qualitative data at scales beyond typical single-researcher projects; but with practices that will be of broad use to scholars thinking through research design and management. Finally, the chapter concludes with reflections on the synergy between longstanding qualitative traditions and evolving digital methods, alongside cautions in the current moment.

## 2. Core Contributions of Qualitative Research on Aging

This section reviews central qualitative approaches and provides examples of their contributions to understanding aging, later life and the life course more generally.

### *2.1 Participant observation and ethnography*

Participant observation is a method that focuses on the direct observation and documentation of human behavior as it unfolds in real-world settings, in real time (Becker, 1958; Gans, 1999; Sánchez-Jankowski, 2002). Sociological ethnography uses participant observation--frequently in tandem with other methods--to examine routines, environments, social structures, and belief systems in broader historical contexts. Ethnography has occupied a central position in the discipline since the early twentieth century (and earlier in adjacent fields), and remains influential in both single-case and comparative sociological studies on a wide range of topics including aging (Abramson, 2021; Abramson & Gong, 2020; Atkinson et al., 2007).

Ethnographic studies often include embedded analyses of observations in social settings (e.g. senior centers or neighborhoods), close attention to individuals' behaviors (e.g. following people as they navigate their days), or a combination of both (Abramson & Sánchez-Jankowski, 2020; Seim, 2024). The granular data this produces provide crucial contributions to the sociology of aging by explaining how mechanisms connect broader social processes like urban inequality to the lives of older people (e.g., Abramson, 2015; Torres, 2025), identifying errors in prevailing theoretical models (Hochschild, 1973), and revealing mechanisms that broader methods overlook (Gubrium, 1976/1997; Myerhoff, 1980). For example, Arlie Hochschild's *The Unexpected Community* (1973) documents how residents of a retirement complex defied then-popular "disengagement theory" by forging strong communal bonds. Myerhoff's *Number Our Days* (1980) showed how Jewish older adults in Venice, California, preserved meaning in later life alongside practical challenges to life and identity. Gubrium's *Living and Dying at Murray Manor* (1976/1997) demonstrated how nursing-home residents and staff mutually shaped daily routines, illuminating structural constraints alongside older adults' agency in revealing the social organization of an institution where many live out their final years. Across these works, participant observation uncovers creative responses to aging and adaptation. By "being there," field researchers captured not only granular data that help illuminate what explanations would anticipate, but also paths through daily life missed or misrepresented in existing accounts (Gans, 1999; Jerolmack & Khan, 2014; Sánchez-Jankowski, 2002).

Participant observation's unique value in aging research is also evident in more recent ethnographies that embed researchers in older adults' daily lives. Torres's *At Home in the City* (2025) exemplifies this by chronicling older New Yorkers in a gentrifying neighborhood. Her long-term immersion in a local bakery--an informal social hub that exists in reality, but would be missed in studies of formal organizations--uncovered how small changes in places of daily gathering can profoundly affect sense of belonging and network strategies. Abramson's *The End Game* (2015) demonstrates how extended fieldwork can link macro-level inequality to observable trajectories in later life, including unpacking puzzles that become clear only with more granular data. Drawing on multi-site observations across four diverse neighborhoods, Abramson documented how health disparities, wealth gaps, and cultural differences converged to shape how older adults handled practical dilemmas ranging from securing medical care to agism. By using qualitative data analysis (QDA) to help compare how older adults from different social backgrounds responded to similar problems, this approach revealed micro-dynamics of stratification to reveal pathways connecting prior inequality, present contexts, and uneven outcomes. Likewise, Buch's *Inequalities of Aging* highlights how home care interactions bind older adults and primarily female caregivers in ways that maintain "independence" but reinforce race, class, and gender inequalities (Buch, 2018). Through in-home observations, they uncovered a paradox of independence that often rests on precarious labor. These cases show that sustained ethnographic work enabled by participant observation can reveal how policy, community, network and economic structures are adapted, enacted, or contested in everyday life.

*2.2 In-Depth interviewing*

In-depth interviewing has become a central method for qualitative and mixed-methods researchers across the social sciences, particularly in the 21^st^ century (Knott et al., 2022). Using semi-structured or open-ended conversations which are typically transcribed and then analyzed as text transcripts, in-depth interviews facilitate understanding older adults' personal histories, behaviors, and beliefs in their own words. Interviews can reveal cultural understandings, as well as beliefs or recollections that might remain hidden if researchers relied exclusively on field observation or structured survey questionnaires (Knott et al., 2022; Pugh, 2013; Weiss, 1995). Further, they allow researchers to understand how people combine challenges, life events, and experiences into narratives that are themselves important for understanding social action (Abramson et al., 2024; Dohan et al., 2016).

In aging research, interviews commonly invite older individuals to discuss topics like family caregiving, health challenges, or financial pressures, allowing the researcher to follow threads of conversation including those that arise spontaneously. For instance, Katherine Newman's (2004) *A Different Shade of Gray* demonstrated how extensive interviews with Black and Latino older adults in inner-city New York offered a window into deteriorating urban conditions, and how inadequate healthcare and minimal public support left them and their families vulnerable. Meika Loe's *Aging Our Way* (2011) chronicled the lives of Americans who lived beyond their 80s, many living at home independently yet engaged with extensive support networks. Her participants described both pride in self-reliance and a steady need for practical and emotional help from family, neighbors, or hired aides. Each picture complicates notions of universalities across backgrounds, binaries of "dependent" vs. "independent" older adults, or successful versus unsuccessful aging, by showing how older adults navigate autonomy and interdependence in their understandings of later life.

Madonna Harrington Meyer's *Grandmothers at Work* (2014) provides another example, with in-depth interviews used to capture role negotiations and pressures in later life (Harrington Meyer, 2014). Drawing on interviews with working grandmothers, she documented how many women in their 50s or 60s postponed retirement or rearranged job schedules to provide childcare, sometimes simultaneously caring for frail spouses or parents, showing how gendered disparities in care work continue to shape later life. Through these personal narratives, the study links structural and historical factors (like limited childcare options, longer lifespans, and later retirement) with the lived realities of older women who juggle paid employment and family obligations, practical challenges and love for family. Notably, all these works engage with broader demographic and survey data, while complementing findings that are important and hard to otherwise capture. Beyond these major works, myriad studies use in-depth interviews to highlight how older adults interpret evolving cultural expectations, social policy gaps, and health vulnerabilities while offering insights into how they understand their own experience.

*2.3 Historical and archival methods*

Historical methods often situate contemporary aging within broader social, political, and media contexts using analysis that are broadly qualitative, and not reducible to numbers. Rather than generating new data in interactions with subjects, historical methods involve analyzing primary and secondary sources like policy documents, archival records, newspapers, extant scholarship and texts, often to see how policies and perceptions operate in a particular moment, or shift over time. Jill Quadagno's *The Transformation of Old Age Security* (1988), for instance, shows how U.S. Social Security emerged from class struggles and interest group compromises, illustrating that decisions made decades ago (such as excluding agricultural and domestic workers) continue to shape many older people's financial security today (Quadagno, 1988). Similarly, James Patterson's *America's Struggle Against Poverty in the Twentieth Century* (2000) situates aging policy within broad cultural designation of "deserving" versus "undeserving" recipients- that has shaped American social policy and the possibility of entitlements used by older adults for generations (Patterson, 2000). Often, historical methods are also integrated into ethnographic or interview studies, such as Klinenberg's *Heat Wave* (2002), which used a variety of data types to provide a 'social autopsy' of the 1995 Chicago heatwave, explaining why it disproportionately killed older adults, and in particular older black men living alone (Klinenberg, 2002).

Some scholars analyze popular portrayals of aging in the media, uncovering how stereotypes, narratives and policy attitudes shape older adults' identities and treatment (Jen et al., 2021). By revealing the historical roots and evolving rhetoric of these frames, content analyses help researchers understand why certain inequalities or misconceptions persist. Ultimately, historical content analysis enriches present-day findings by adding a temporal dimension. It clarifies that older adults' current experiences--documented via participant observation or interviews--are partly shaped by long-running policy histories, economic shifts, and cultural discourses that stretch back decades or more.

*2.4 Why qualitative work matters for aging*

In the examples above, qualitative approaches--understood broadly in pragmatic terms as data sources not reducible to numbers or quantitative models alone (Abramson et al., 2026)--offer distinctive insights into older adults' day-to-day routines, decision-making, and structural constraints. Consider understandings of cumulative advantage and disadvantage--i.e., an empirically supported model of how inequalities accumulate over the life course, well charted in quantitative inquiries (Dannefer, 2003; Falletta & Dannefer, 2014; Jackson & Engelman, 2022). Demographic and epidemiological data show central trends in morbidity, mortality, and life expectancy. Ethnographic and interview-based studies capture how these core inequalities operate in concrete settings, such as "senior centers" or home-care agencies, and how older adults develop adaptive strategies in the face of limited resources-- both concretizing and adding granularity to our knowledgebase (Abramson, 2015; Torres, 2025). Historical analyses, meanwhile, reveal how policy contexts and media narratives have shaped opportunities and constraints available to older adults, often in ways that exacerbate vulnerability (Harrington Meyer, 2014; Klinenberg, 2002; Quadagno, 1988).

*2.5 Limitations of traditional qualitative approaches*

Qualitative approaches--particularly single-site studies which focus on researcher experiences--have been critiqued for limited generalizability, transparency and replicability even as they are recognized for enabling insights for decades (Goldthorpe, 2000; Small & Calarco, 2022). Work on aging has not been immune. The continuing integration of qualitative data has been positioned as a way to advance understanding granular mechanisms, analytical processes, and experiences while advancing theory (Alley et al., 2010; Abramson & Portacolone, 2017; Phoenix 2018). Yet, concerns about the extent to which findings are generalizable, systematic, and scientific remain, as in other arenas of social science (Abramson, 2021). An added source of confusion is that the research paradigms used in qualitative research vary widely (Gong & Abramson, 2020a; Mays & Pope, 2000; Rendle et al., 2019). The next sections examine how new digital tools, larger datasets, and collaborative strategies are expanding the scope of qualitative aging research. These developments address core critiques of traditional qualitative inquiry while maintaining the depth that has been a hallmark of qualitative contributions and acknowledging the challenges of contemporary qualitative analysis.

## 3. Changing Technologies, Computational Analysis and "Big Qualitative" Data

Developments in social science offer possibilities for those seeking more verifiable and reproducible approaches to qualitative research, now seeing broader use across the social sciences as well as in aging research. New technologies that are used for large scale, typically funded initiatives, are also making their way into smaller projects--either out of convenience, or requirements for data sharing and transparency from institutions such as the National Institutes of Health (NIH, 2023).

*3.1 Changing context and technology*

Recent decades have seen an expansion of multi-site team-based qualitative research, accompanied by groundwork for more robust sampling protocols, data sharing repositories, verification initiatives, and expanding software tools for systematic analysis (DeLuca et al., 2016; Dohan & Sánchez-Jankowski, 1998; Edin et al., 2024; Murphy et al., 2021). Analytical possibilities are being radically revisited and contested as artificial intelligence enables the automation of core qualitative practice, such as coding/indexing text, summarization, and retrieval--for better or for worse, this is increasingly being integrated into tools ranging from literature searches, to search algorithms, to dedicated qualitative software, and even to spelling/grammar checks. These developments have broad effects on society and research across varied domains (Abramson et al., 2026; Fourcade & Healy, 2024; Nguyen-Trung, 2025).

Digital transcription technology, online data repositories, institutional mandates for data sharing, large-scale mixed-methods and interview initiatives including qualitative data (such as the American Voices project or National Dementia Workforce Study), the maturation of machine learning in computational text analysis, and the proliferation of large language models have all contributed to what some are calling the rise of "big qualitative data." Such an approach generates data using conventional methods like in-depth interviews and participant observation, but at a much larger scale than seen previously (Abramson et al., 2024; Chandrasekar et al., 2024; Nelson, 2020; Murphy et al., 2021).

In parallel, tools from computational social science are being used in individual and team-based qualitative studies to extend the raw capacity to record interactions as text, index that text as data, code store and retrieve individual examples or thousand page excerpts, and reuse data for multiple research questions alongside calls to "scale up" inquiry and enable an open-science framework (Abramson et al., 2018; Bernstein & Dohan, 2020; DeLuca et al., 2016; Edin et al., 2024; Than et al., 2025). New methods for bridging quantitative text analytics with interpretive close readings--sometimes described as "scaling down" big data to re-embed meaning--offers compelling pathways to look at clusters of themes, topics and behavioral sequences in text data (Breiger, 2015; Nelson, 2020; Pardo-Guerra & Pahwa, 2022). Work to expand crowdsourcing and large-scale collaborative undertakings have parallels to initiatives like Mass Observation in the 1930s, but with modern tools, methods and a deliberate aim to test how carefully organized qualitative data might be used by research teams to replicate analytical claims and tie findings to validated quantitative measures (Edin et al., 2024).

### *3.2 Methodological tensions*

Yet as methodological tools and data sets expand, scholars may encounter tensions. Qualitative inquiry encompasses a range of epistemic positions--from exploratory and comparative approaches to extensions of conventional science and humanistic critique (Gong & Abramson, 2020b; Rendle et al., 2019; Small & Calarco, 2022). Some traditions explicitly reject objectivity or replication (Burawoy, 1998; Glaser & Strauss, 1967; Collins, 2000), while others view qualitative work as complementary to science (Lamont & White, 2009; Small & Calarco, 2022) or fundamental to it (Abramson & Sánchez-Jankowski, 2020; Gans, 1999; Sánchez-Jankowski, 2002). This diversity generates misunderstandings. In aging research, where qualitative methods have a less established history, the problem is acute: systematic comparative work may be labeled "exploratory" (Abramson, 2021), and humanist scholarship evaluated using scientific standards.

Methodological pluralism--the deliberate integration of diverse approaches to match the complexity of sociological phenomena--provides one avenue to address this tension (Lamont & White, 2009). Every method carries hazards of omission; multiple techniques can illuminate different facets of phenomena, while enabling triangulation and complementary insights within and across projects (Abramson & Gong, 2020; Cicourel, 1964; Lamont & Swidler, 2014; Small, 2011). Yet critics caution that large-scale qualitative data operate under a logic consistent with explanatory research within a broadly scientific paradigm, that may not align with humanist approaches where depth rather than scale constitutes the source of legitimacy (Abramson et al., 2018; Lareau & Rao, 2016). Further, qualitative work done at scale with deep granular information, represents a distinct and emergent research "genre": scope adds new strengths (e.g., generalization) alongside challenges (e.g., quality control, text processing) closer to computational analyses of 'big data'--straining direct analogy to simplified conceptualization of traditional qualitative methods, which in practice have long varied in dispositions toward conventional science (Abramson et al., 2026).

### *3.3 Value added of integration*

For scholars of aging, linking micro-level insights to broader demographic and epidemiological patterns over the life course is both a standalone contribution and a potentially powerful way to bridge otherwise disparate literatures (Abramson, 2021). For instance, connecting aggregate patterns of advantage and disadvantage over the life course to observations in institutional settings reveals the mechanisms that shape trajectories for older adults and relates them to general conceptual models. Historically, the sociological canon has embraced pluralism in the works of foundational figures such as Du Bois, Durkheim, and Wells, who drew on multiple data types and analytical styles before "mixed methods" was a recognized term. Today, multi-method designs that integrate computational tools with large-scale qualitative data offer new insights positioned between population-level processes and the granular interpersonal details of everyday life. Such approaches retain the hallmark openness to emergent findings in qualitative research while incorporating the scale, reproducibility, and heterogeneity that computational analyses of larger projects provide (Li & Abramson, 2024; Nelson, 2020; Pardo-Guerra & Pahwa, 2022). Extending traditions of sociological pragmatism and pluralism, these "new hybrid" methods complement rather than supplant interpretive practices in qualitative aging research. They add a research genre capable of broader comparison and pattern discovery--and perhaps of building archives of present moments for the future, as envisioned by Du Bois and Mills (Abramson, 2024).

## 4. Case Studies: Comparative Ethnography and Secondary Analysis

To illustrate how large-scale qualitative data combined with advanced computational analysis can deepen understandings of aging and the life course, I turn to several ongoing or recently completed projects. The first example focuses on a team ethnography supported by computational tools. The second highlights a secondary analysis of narratives from a large-scale, nationally representative interview project. In each case, I draw parallels to other studies relevant to aging, note ongoing developments in the field, and examine strategies that may be valuable more broadly for qualitative research in an increasingly computational moment.

### *4.1 Case 1: Team ethnography*

In studies of aging, "team ethnography" projects--where multiple researchers collaborate across sites, aided by digital tools--have gained visibility, particularly after the National Institute of Health's 2011 best practices report recognized primary qualitative data as a valued source (Creswell et al., 2011). Early dementia care studies (Scales et al., 2008) showed that collaboration is not simply a matter of adding more researchers, but a strategy for gaining greater perspective, coverage, and opportunities for triangulation. Building on sociological and anthropological archives, these efforts underscore that sharing data, interpreting findings across diverse backgrounds, and fostering trust among collaborators and communities are not merely technical tasks but epistemic and practical commitments (Abramson & Dohan, 2015). In aging research, this is especially critical for conditions such as Alzheimer's Disease and related dementias (ADRD), where sensitivities around patient identity, family involvement, and institutional boundaries demand careful protocols.

Team strategies can capture multiple vantage points, distribute fieldwork, and strengthen confidence in shared findings beyond what single-researcher models allow (Bernstein & Dohan, 2020). At the same time, collaboration entails added coordination and cost, and technology cannot replace the thoughtful design, interpersonal relationships, time, and energy needed to generate high-quality ethnographic data on aging (Abramson, 2021). Still, a growing array of tools now augments workflows such as file collection, audio transcription, fieldnote naming, subject deidentification, indexing, pattern analysis, and revisiting prior findings (Abramson et al., 2026; Li & Abramson, 2024). Used alongside face-to-face field engagement, these tools extend prospective qualitative studies by streamlining and broadening primary data collection while creating new opportunities for secondary analysis in aging research (Abramson et al., 2026).

#### *4.1.1 DISCERN*

The DISCERN study offers an example in this prospective sense--integrating computational methods with comparative ethnography as a design decision. The use of computation was deliberate and used to facilitate rather than replace human insight into the complex experiences and challenges of life with dementia, and all materials remain accessible without computational expertise (Arteaga et al., 2025).

DISCERN is the result of a multi-year NIH funded project examining how older adults--particularly those diagnosed with Alzheimer's Disease and Related Dementias (ADRD)--navigate complex treatment decisions alongside their family care partners. This work builds on lessons from prior multimethod inquiries in the study of later life, including the PtDelib study of advanced cancer patient trajectories encompassing nearly 300 multi-wave, in-depth interviews, validated survey metrics, more than 5 years of field observations in multiple states, and hand coding and computational analysis of more than 12,000 pages of text (Abramson & Dohan, 2015; Garrett et al., 2019; Li et al., 2021).

To understand diverse ADRD experiences across 3 states, we drew on multiple data types and pilot observations to identify sites, then conducted sustained observations of people and organizational settings over time (Small & Calarco, 2022). Led by Daniel Dohan at the UCSF Medical Cultures Lab, the project also employs respondent driven sampling to recruit both family caregivers and organizational actors in social networks, situating direct observations within wider contexts of later life. Throughout the study, we used a multimethod analytic approach that connects qualitative data analysis software to systematic indexing, cleaning, and text organization using both conventional qualitative data analysis and tools from machine learning and natural language processing (Abramson et al., 2024; Deterding & Waters, 2021; Li & Abramson, 2025). Standardized labeling and organization--consistent file naming and shared subject IDs--enabled linking across interpretive close readings, dictionary-based indexing, and supervised machine learning (discussed below).

As in prior projects, the team met regularly to discuss findings, challenges, and analytic categories. This was guided by project aims, research questions and extant literature, but also reflected patterns that emerged in the data and fieldwork--following a contemporary iterative approach seen in both qualitative sociology and computational social science (Li & Abramson, 2025; Deterding & Waters, 2021). Analyses were neither strictly deductive nor inductive, but reflected prior concerns based in research and policy needs, as well as unexpected results as data continued to accumulate throughout the project. To provide examples, relevant issues such as "migration_history" or "social_isolation" were initially identified by humans as related to the study and seen in our data in discussions, then operationalized using dictionaries (i.e., word lists) or machine-learning classifiers trained on human-coded examples (i.e. small local language models (Li et al., 2021; Mostafavi et al., 2025) to index text with relevant concepts in the context of larger interview or field documents (Li et al., 2021; Lichtenstein & Rucks-Ahidiana, 2023; Mostafavi et al., 2025). After the automated process, human review and triangulation verify or adjust the machine's output--weeding out errors in a way that keeps human domain knowledge central in the "tagging" of qualitative data (Abramson et al., 2026).

Yet, initial passes to identify relevant segments from close to 10,000 pages of data do not replace or displace in-depth human reading. Field reflections, deeper interpretive reading of text segments in computer assisted qualitative data analysis software that show full context (we used ATLAS.ti, but this would work in other platforms), and memo writing illuminate how findings unfold--augmented by prior taggings like "migration_history," yet never reduced to it. Even early benchmark data suggested that hybrid coding can be both reliable and efficient, while also identifying disconfirming evidence or outliers humans missed (Li et al., 2021; Nelson et al., 2021).

DISCERN does not treat algorithmic outputs as self-sufficient, and like the AVP, researchers use the data for different types of projects-- traditional small-n analyses alongside computational and mixed-method work. However, in each variant, trained ethnographers and interviewers always return to transcripts, field notes, and memos--alongside computational results--to interpret findings. This was not done in an exploratory manner focused on induction (as in traditional grounded theory), but in light of existing questions, epidemiological patterns, and patterns in the lives of older adults and carepartners that included both expected outcomes and novel findings. DISCERN notably demonstrates how computational text analysis can help enable analyses within and across multiple sites, and highlight patterns in new ways such as through a pairing of case analyses and ethno-array heatmaps (Abramson & Dohan, 2015; Arteaga et al., 2025). Yet, the use of computation preserves the micro-level text accounts essential qualitative analysis and writing. By illuminating how history, local worlds, and individual biographies intersect at critical life-course moments, the project shows that rigorous research located between single-site ethnographies and massive qualitative platforms can purposefully integrate new technologies to contribute to knowledge and policy.

The callout below reviews some of the study practices and presents suggestions that may be applicable for future studies.

**Useful Practices and their Logics (Ethnographic)**

**The following practices from our work on DISCERN are important to team-based workflow but have use more broadly. Some have been central to my own individual projects on aging going back to my time a solo ethnographer in graduate school (Abramson, 2011, 2015).**

**File Naming**

**A shared file-naming convention encodes metadata directly into the filename, which facilitates later retrieval and analysis. This structure allows files to be sorted or grouped in qualitative software and enables researchers to treat filename segments as variables in a *computational data frame* (essentially, a table or spreadsheet for analysis).**

**While valuable for small-scale studies, this practice is also important to large qualitative inquiries.**

**In my standard practice, a filename like WCadc_20230504_INTV_OA010_CA.mp3 is structured as [Site]_[Date]_ [Data Type]_[Subject ID]_[Researcher Initials].**

**This system allows one to quickly group all data by site, date, or researcher. After transcription, the resulting .txt file maintains this structure, ensuring consistency across the analytical workflow (Li & Abramson, 2024).**

**Transcription Norms**

**Transcripts should be formatted with a consistent speaker. For instance, I often advise IDs that include alphanumeric indicators for type of respondent and when they enter the study (e.g., OA010 for an older adult, INTV_CA for the researcher, which may get filtered out in computational models). This allows for linking data across different file types and prepares the text for computational analysis. All the data with OA010 can be linked with either a coding language (e.g. python/R) or in QDA software (e.g. ATLAS.ti, NVivo).**

**Consider this example from my first book on aging (Abramson, 2015):**

***INTV_CA: And so, can you tell me a little bit about retirement?...***

***OA017: It's, growing older is weird... My body is 73, but my mind is about 35. I don't know how I'm supposed to act as a senior...***

**The speaker IDs (INTV_CA, OA017) index the text for export to machine learning applications. Using a consistent format and saving files with UTF-8 encoding minimizes compatibility issues with both machine learning workflows and qualitative software.**

**Fieldnotes and Memos**

**How ethnographers produce fieldnotes varies, but for team-based or computational analysis, a few principles are key (Abramson & Gong, 2020; Li & Abramson, 2024).**

**1.** Use Subject IDs: **Including IDs (e.g., OA017) in fieldnotes allows for linking observations of an individual across different contexts, either through qualitative software or simple text searches.**

**2.** Purposeful Formatting: **Use paragraphs and line breaks to group meaningful interactions or "chunks" of action. This initial structuring aids in identifying patterns later.**

**3.** Distinguish Observation from higher Level Interpretation: **Using brackets, like [[...]], to denote analytical thoughts or memos helps team members distinguish the ethnographer's interpretation from direct observation.**

**For example, a fieldnote segment from *The End Game* (edited) reads:**

***Went to see OA021. She said she felt well, asked about my mom...She said her family was fine and visited "yesterday" as always...OA021 did not know what she was going to do for easter but said "I hope they have services here. They always do. It is a nice facility [[I was surprised she used this word]]...***

**While some teams use pseudonyms from the start, I prefer using IDs during the analytical phase because they scale efficiently for connecting data. Pseudonyms can then be assigned during the final writing process to re-humanize the data for the reader while protecting subject confidentiality.**

***Storage and Analysis***

**Our analytic workflow rests on three principles: a secure and coherent data infrastructure, reproducible analysis, and iterative inquiry.**

**1.** Data Infrastructure: **In practice, transcripts and fieldnotes are stored in HIPAA- or human-subjects--approved cloud systems, with each file labeled to align with site or participant descriptors. Identifying information is maintained separately in a key linking contact data to codes in a database. A spreadsheet may contain identifiers alongside demographic or survey measures to serve as metadata for qualitative software or to structure tabulated text for computational analysis (Li & Abramson, 2024). Contemporary qualitative programs allow text to be interactively organized, exported for quantitative or computational analysis, and then reimported as needed (Abramson, 2022ab).**

**2.** Reproducible Analysis: **Analyses are documented in computational notebooks (e.g., Jupyter for Python, Quarto for R) or scripts, alongside analytic memos that provide narrative accounts of sites and cases. Crucially, any cleaning of text occurs only in analytic notebooks, leaving raw data files unaltered. This scaffolding mirrors an open-science ethos: it ensures transparency, allows replication by team members or collaborators, and facilitates tools such as named entity recognition to blind identifiable data before publication.**

**3.** Iterative Inquiry: **The research process is recursive, consistent with ethnographic logics and contemporary iterative approaches to indexing and coding (Abramson et al., 2018; Deterding & Waters, 2021; Lichtenstein & Rucks-Ahidiana, 2023). Pilot analyses often surface unanticipated themes (e.g., new economic arrangements), prompting adjustments to coding or additional observations. Computational models may reveal latent biases--such as over-reliance on medicalized accounts of aging--guiding deeper interpretive readings (Abramson, 2024; Nelson, 2020). By systematically naming and formatting text, teams can query corpora for contextualized references (e.g., diagnosis) and develop memos as bridges to scholarly writing. Automations--dictionaries, key words, and supervised machine learning (Deterding &**

**Waters, 2021; Li et al., 2021)--pool relevant material efficiently while keeping interpretive depth central. This can be done in pragmatic ways, with some topics using automation, and others, reserved for in-depth human reading (Abramson et al., 2026).**

*4.2 Case 2: Secondary data analysis*

While DISCERN demonstrates the immediate applicability of this workflow to prospective projects, the same logic extends retroactively in "big qualitative data" initiatives like the AVP where scholars can perform secondary analyses of data gathered qualitatively.

*4.2.1 The American Voices Project (AVP)*

The American Voices Project (AVP) is a nationally representative study that contains thousands of in-depth interviews, supplemented by surveys, from participants across all 50 states. Its design examines Americans' everyday lives and experiences, covering issues central to aging research, such as caregiving, health transitions, family dynamics, and economic vulnerability. A centerpiece of the AVP is its open-science ethos: transcripts are de-identified and made available to researchers outside the data collection team, who can conduct new analyses or replicate findings in ways more often found with survey data or archival research. Teams of scholars have used AVP data in varied projects that combine traditional qualitative approaches, computational text analysis, and/or merge computational text analysis with traditional interpretive strategies (for instance, visualizing patterns using natural language processing, then offering in-depth narratives). Several papers have had their core claims verified or replicated in accordance with the data set's science principles, and workshops allowed scholars to cross-check both work and code. Data are geographically coded (including in the example below), and include oversampling with weights to preserve representativeness when analyzing heterogeneous populations (Edin et al., 2024).

Extant studies use segments of the data for traditional qualitative data analysis on topics including aging. In the initial wave of papers, Freitag (2024), used a subset of 25 in-depth interviews with low-income individuals over age 50 receiving Supplemental Security Income (SSI). Their analysis revealed that while SSI often functions as an income source of last resort, the benefit level is too meager to ensure independence for many in later life. Older adults described the challenges of meeting monthly costs and noted that strict SSI rules penalize informal or family-based support, which frequently undermines recipients' efforts to cope. Freitag's work demonstrates one way that AVP data can be broken into smaller subsets for highly focused "traditional" qualitative analysis, adding depth to broader statistical findings on topics like precarity and unequal aging (Carr, 2019). These approaches benefit from the dataset's broad sampling frame and ability to capture older adults during historical moments such as the initial wave of the COVID pandemic.

Researchers have also leveraged the entirety of AVP for analyses of interest to aging and the life course. Abramson et al. (2024) conducted a study of chronic pain--an issue spanning age groups yet disproportionately affecting older adults--by applying dictionary-based indexing of pain-discussions and then expanding with machine-learning to identify pain-related narratives across more than 1500 interviews. Our team then modeled language patterns using semantic network analysis to make visible patterned variation, and linked these patterns to discussions of pain management, stigma, and moral framings of narcotic use which shared both commonalities and differences by education and gender--tying into broader challenges of navigating pain across the life course, and in later life. Many of the topics reflect ongoing fieldwork in projects like DISCERN, and issues touched upon ethnographically in discussions of pain and stigma among older adults in *The End Game* (Abramson, 2015).

Lessons from projects like AVP are important for scholars of aging, in part because new initiatives aim to produce large-scale qualitative data on aging and lifecourse processes. In the arena of policy and public engagement, The People Say focuses on sharing the beliefs and experiences of older adults on a variety of topics, and connecting to a user friendly web interface (Mauldin et al., 2024). Notably, the National Dementia Workforce Study (NDWS) includes expansions from an ambitious survey effort to a large-scale scientific endeavor including a multi-method format that will provide unparalleled scope for in-depth interviews charting experiences and decisions in the domain of care work for older adults living with dementia (Maust et al., 2025). Such data sets and new generations of scholarship are likely to be supported by computational methods, which increasingly help aggregate, process, analyze, and curate qualitative data across a range of topical areas (Abramson et al., 2026; Than et al., 2025).

Other emerging work likewise leverages secondary qualitative data. For instance, Hacking et al. (2022) combines computational text analysis with human readings to identify sentiments associated with understanding the 'nursing home' as a social context (Hacking et al., 2022). Such works confirm that text mining and AI-driven classification can expedite the handling of large volumes of staff notes, logs, and other forms of text, potentially unveiling patterns in resident interactions, treatment regimens, or staff communication that might otherwise remain hidden--or at minimum, help reveal aggregate patterns to situate concrete case examples as typical or disconfirming (Small & Calarco, 2022). These technologies can flag high-level patterns, such as which groups espouse positive or negative beliefs about various facets of later life, or reveal themes of interest. At the same time, scaling down to see how social forces operate in everyday life remains crucial to connect to traditionally qualitative narratives and avoid error (Carlsen & Ralund, 2022). Beyond these cases, computational analyses of historical documents, media, and online forums provide further avenues for studying how representations of aging, caregiving, and later life shift over time, building on now classical work in historical and computational sociology (Mohr & Duquenne, 1997).

To illustrate how large-scale data collection and analytic strategies involving computational and qualitative analysis can potentially contribute to research on aging, I revisit some practices from my team's analysis of the over 1500 in depth interviews in the AVP.

**Useful Practices and their Logics (Secondary Analysis of Interviews)**

The following example, from a secondary analysis of the American Voices Project, illustrates how a mixed-method design can connect large-scale patterns with in-depth narrative detail (Abramson et al., 2024). Although this example focuses on pain, the practices are directly applicable to studying complex topics central to aging research, such as how older adults narrate their experiences with health, loss, or social connection--all topics of recent works, following similar workflow.

1. **Sub setting:** We began with a deep reading of interviews to understand the narrative landscape and familiarize ourselves with the data. We then used a Word2Vec model, a type of local language model that can be run on a consumer-grade hardware without uploading to a commercial service, to model language on topics of interest represented in the text. This approach integrated both prior interests and emergent readings, but moved beyond keyword searches to capture varied and nuanced ways respondents discussed their experiences.
2. **Visualizing Narrative Structures:** We used semantic network analysis to visualize patterns in language . These networks revealed frequently co-occurring words and thematic clusters (e.g., "medical management," "broken body"), as well as if and how these varied by groups. This allowed us to map the general patterns in how people make sense of their suffering and, crucially, to see how these schemas varied across different demographic groups.
3. **Connecting Macro Patterns to Micro Narratives:** The core of the practice is the iterative movement between the macroscopic view from networks and the microscopic view of individual narratives. The large-scale patterns helped us identify particularly revealing passages for in-depth reading. In turn, the close analysis of these transcripts explained *why* respondents in different life circumstances, such as older adults facing distinct health challenges, described their experiences so differently. This process links broad patterns to the detailed, accounts that produce them--and works well for field notes, interviews, and even related documents in our analyses with DISCERN and other aging specific studies.

## 5. Conclusion

Participant observation, in-depth interviews, and historical inquiry have long contributed to social scientific understandings of aging--shaping social scientific accounts of community, inequality, institutions, and culture (Myerhoff, 1980; Harrington Meyer, 2014; Gubrium, 1997; Hochschild, 1973; Quadagno, 1988; Newman, 2004). This chapter has argued that computational social science (CSS) tools--such as text analysis, machine learning, and natural language processing (NLP)--offer powerful potential for extending and expanding qualitative inquiry. When used purposefully by individuals and teams, the current computational repertoire can aid in streamlining workflow, allowing new mixed-methods analyses, and supporting large-scale endeavors that combine systematic sampling with deeply contextual information. This expands possibilities for understanding myriad facets of aging, through team ethnographies, large-scale in-depth interview studies, and secondary analyses of historical and qualitative data in growing repositories.

This chapter offered concrete examples from initiatives like DISCERN (a comparative ethnography of life with dementia), the American Voices Project (a national-scale interview study enabling secondary analyses of later life) as well as emergent expansions. Examples show how detailed observations and interview responses can be paired with computational strategies to locate, filter, and cluster large volumes of text, then link back to close reading in ways that preserve local meanings, while facilitating explanation and comparison on topics ranging from pain to precarity in our later years (Abramson et al., 2024). Projects such as DISCERN illustrate how systematic data management, collaboration, iterative indexing of text, and shared reference systems help multi-researcher teams coordinate observations on complex topics like dementia care decisions while maintaining confidence that findings are not an artifact of a single researcher or site. Such "team ethnography" requires close integration of computational infrastructure and human review. This synthesis is essential for combining the grounded perspective provided by field notes and extended interviews, verifying that older adults' narratives are neither diluted nor decontextualized, and ensuring the data are usable for traditional narrative analyses as well as large-scale comparison (Abramson et al., 2026; Arteaga et al., 2025). Some of the lessons and practices shared here may facilitate new forms of inquiry, or enhance existing work, in ways that can contribute to understanding aging and the life course, and the sections above identify which practices are transposable outside of team-based studies.

Likewise, the discussion of the AVP study demonstrates the possibilities of nationwide reach and open-science principles, revealing how the same study can support focused analyses on older adults (Freitag et al., 2024) as well as full-corpus studies linking pain, health, and inequality across the life course (Edin et al., 2024; Abramson et al., 2024; Mauldin et al., 2024). Studies of this sort are likely to migrate to aging and other domains of social inquiry, and the examples in this chapter show how computational tools can aid in streamlining qualitative analysis (data organization and indexing), scaling (deidentification and aggregation of data), and creating mixed-method analyses for relevant topics (computational text analysis with close reading). These studies also point to the continued importance of reflexivity about how we both collect and make sense of data. When used poorly, any research tool can be dangerous, particularly when misunderstood or overhyped, as is often the case with AI (Abramson et al., 2026).

A review of the connection between new technologies and society is beyond the scope of this study, but the perennial risks of error and misrepresentation do not dissolve with either scale or computation. Risks include "precisely inaccurate" findings from "big data" not carefully contextualized or read, uncritical use of language models with embedded biases about older adults and marginalized groups, ignoring disconfirming findings to focus on favored narratives, and elevating computational instruments above human insights and the goal of understanding what it means to age (McFarland & McFarland, 2015; Farber et al., 2025; Carlsen & Ralund, 2022; Li & Abramson, 2024). Algorithmic bias poses risks in aging research. Automated text analysis can reproduce ageist assumptions embedded in training data, reinforcing harmful stereotypes about older adults' cognitive abilities or social engagement. Without careful human oversight, machine learning models might misclassify discussions of memory concerns as dementia symptoms or interpret limited socializing with peers--which can be an adaptive strategy for managing energy in later life--as pathological withdrawal. These algorithmic errors can systematically distort research findings about aging experiences. Such concerns are not new but take on sharper stakes in an era of widespread AI integration, where chatbots offer to analyze data in our stead--and when used poorly, risk magnifying human biases rather than offering insight. At the same time, a growing body of work shows how new technologies can be used to reduce errors and broaden perspectives, particularly when used on narrow tasks in a thoughtful way (Abramson et al., 2026; Than et al., 2025).

In the current moment--as an aging population converges with technological and social unsettlement--sociological frameworks, skills, and lenses are critical to advancing our understanding of aging and later life. Computational tools, when leveraged thoughtfully, expand the methodological repertoire of qualitative aging research in at least three ways. First, they facilitate more comprehensive coverage, reaching diverse older people--rural, urban, or institutionalized--whose experiences might otherwise remain unobserved in the context of an otherwise accurate and coherent study. Second, they allow more robust validation and verification, helping address concerns about qualitative verification, while allowing enough data for meaningful deidentification and qualitative data sharing to reproduce analyses (Edin et al., 2024; Abramson & Dohan, 2015). Third, they enable flexible, multi-level insights, revealing not only micro-level dynamics, but macro-patterns, and the sociohistorical contexts in which older people live their lives--in ways that do not replace quantitative or qualitative approaches, but allow connecting them empirically in new ways. The ability to connect longer sequences of data, link text, and triangulate findings within and across teams allows new tests of claims--for example, whether and how accumulated advantages and disadvantages converge to shape later-life possibilities.

Tools for modeling language and connecting to narrative might also shed light on how older individuals within and across societies make sense of important issues like bereavement, economic downturns, pandemics, and even technology in ways not yet fully understood. International collaborations can enrich comparative work, as possibilities for aggregating behaviors observed in situ can chart routines across time and space while connecting to a local vantage (Bernstein & Dohan, 2020). Such developments illustrate the potential—of *streamlining*, *scaling*, hybrid *mixed-method analyses*, and reflexivity—to contribute to understanding aging. However, each does so by broadening--rather than replacing--the methodological foundations of qualitative research, which has produced irreducible insights into later life.